\documentclass[11pt, a4paper, reqno]{amsart}
\usepackage{amsthm, amsmath, amsfonts, amssymb, appendix, dsfont, latexsym}
\usepackage{amsfonts, mathrsfs}
\usepackage{graphicx}
\usepackage{tikz}
\usetikzlibrary{arrows}

\makeatletter
\usepackage{delarray, a4, color}
\usepackage{euscript}
\usepackage[latin1]{inputenc}
\usepackage{enumitem}
\usepackage{hyperref}
\hypersetup{
   colorlinks,
   menucolor=black,
   linkcolor=black,
   citecolor=black,
   urlcolor=blue
}

\definecolor{darkgreen}{rgb}{0,0.5,0}
\definecolor{darkred}{cmyk}{0,1,1,0.4}

\theoremstyle{plain}
\newtheorem{theorem}{Theorem}
\newtheorem*{theorem*}{Theorem}

\theoremstyle{definition}

\theoremstyle{remark}

\newtheorem{remark*}[theorem]{Remark\textup{*}}

\DeclareMathAlphabet{\mathpzc}{OT1}{pzc}{m}{it}

\def\C {\mathbb{C}}

\def\N {\mathbb{N}}

\def\R {\mathbb{R}}

\def\Z {\mathbb{Z}}

\newcommand{\keyword}[1]{\textbf{#1}}

\newcommand{\bA}{\mathbf{A}}
\newcommand{\bB}{\mathbf{B}}

\newcommand{\bJ}{\mathbf{J}}

\newcommand{\bx}{\mathbf{x}}
\newcommand{\by}{\mathbf{y}}

\newcommand{\cC}{\mathcal{C}}

\newcommand{\cE}{\mathcal{E}}

\newcommand{\hH}{\hat{H}}
\newcommand{\hT}{\hat{T}}

\newcommand{\sSU}{\textup{SU}}

\newcommand{\sU}{\textup{U}}

\newcommand{\sx}{\textup{x}}

\newcommand{\CLGN}{C_{\rm LGN}}

\newcommand{\cETF}{\cE^{\mathrm{TF}}}

\newcommand{\CTF}{C^{\mathrm{TF}}}
\newcommand{\CLT}{C^{\mathrm{LT}}}

\newcommand{\NLL}{\mathrm{NLL}}

\DeclareMathOperator{\infspec}{\mathrm{inf\, spec\,}}
\DeclareMathOperator{\curl}{\mathrm{curl}}

\newcommand{\norm}[1]{\left\|#1\right\|}

\newcommand{\sym}{\mathrm{sym}}

\newcommand{\loc}{\mathrm{loc}}

\newcommand{\bDelta}{{\mbox{$\triangle$}\hspace{-8.0pt}\scalebox{0.8}{$\triangle$}}}

\usepackage{mathtools}

\usepackage{placeins}

\title{2D magnetic stability}

\author[D. Lundholm]{Douglas LUNDHOLM}
\address{Department of Mathematics, Uppsala University, Box 480, SE-751 06, Uppsala, Sweden}
\email{\url{douglas.lundholm@math.uu.se}}


\begin{document}

\begin{abstract}
This article is a contribution to the proceedings of 
the 33rd/35th International Colloquium on Group Theoretical Methods in Physics (ICGTMP, Group33/35)
held in Cotonou, Benin, July 15-19, 2024.
The stability of matter is an old and mathematically difficult problem, 
relying both on the uncertainty principle of quantum mechanics and on the 
exclusion principle of quantum statistics. 
We consider here the stability of the self-interacting almost-bosonic anyon gas, 
generalizing the Gross--Pitaevskii / nonlinear Schr\"odinger energy functionals 
to include magnetic self interactions. 
We show that there is a type of supersymmetry in the model which holds only 
for higher values of the magnetic coupling but is broken for lower values, 
and that in the former case supersymmetric ground states exist precisely at 
even-integer quantized values of the coupling. 
These states constitute a manifold of explicit solitonic vortex solutions whose 
densities solve a generalized Liouville equation, and can be regarded as 
nonlinear generalizations of Landau levels. 
The reported work is joint with Alireza Ataei and Dinh-Thi Nguyen 
and makes an earlier 
analysis of self-dual abelian Chern--Simons--Higgs theory by Jackiw and Pi, 
Hagen, and others, mathematically rigorous.
\end{abstract}

\maketitle

\section{Introduction}\label{sec:intro}

\keyword{Quantum statistics} usually refers to the exchange symmetry 
implemented 
by a multi-component quantum state, 
in the case that the components are identical and interpreted as
indistinguishable particles.
In three-dimensional space, the relevant exchange group is the permutation group
which has only two 
one-dimensional representations:
the trivial one, respectively, the sign of a permutation,
resulting in a dichotomy of the ``fundamental'' quantum particles into 
\keyword{bosons} repectively \keyword{fermions}.
Their distinction is most clearly understood by the lack or presence 
of the Pauli exclusion principle,
which actually sets the stage for (arguably) most
of our observed physical reality,
including the coherent gravitational and electromagnetic fields 
(gravitons and photons are bosons),
correlated and possibly entangled laser beams, 
as well as condensed quantum gases, 
respectively properties of materials such as conduction and reflectivity 
(electrons are fermions, that ripple a Fermi sea),
as well as the extensivity and stability of matter up to the scale of planets and stars.
However, about half a century after the discovery of these
Bose--Einstein respectively Fermi--Dirac statistics
--- and now, coincidentally, about half a century ago ---
it was realized that these concepts also allow for generalization
to the braid group in two spatial dimensions 
(and in one dimension if the orientation of exchanges can be consistently tracked), 
with corresponding identical particles
now known as \keyword{``anyons''} 
\cite{LeiMyr-77,GolMenSha-80,GolMenSha-81,Wilczek-82a,Wilczek-82b,GolMenSha-85}. 
Understanding the precise relationship between exchange and exclusion 
for a many-particle gas of such anyons is a difficult fundamental problem 
in mathematical physics, and
we will here report on some recent progress on this problem 
(expanded upon in \cite{Lundholm-23,AtaLunNgu-24}),
which exhibits a non-trivial interpolation 
between the extremal cases of bosons and fermions.

\smallskip

Recall that a \keyword{Schr\"odinger wave function} $\Psi$ for $N$ particles in $\R^d$ is 
conventionally regarded as an $L^2$ (square-integrable, and normalized) function
$$
	\Psi\colon (\R^d)^N \to \C, \qquad \sx = (\bx_1,\ldots,\bx_N) \mapsto \Psi(\sx),
$$
and encodes a probability distribution on the configuration space $(\R^d)^N$
of \emph{distinguishable} particles:
$$
	|\Psi|^2\colon (\R^d)^N \to \R_+, \qquad \sx \mapsto |\Psi(\sx)|^2.
$$
\emph{Indistinguishable} particles necessarily must have permutation-symmetric distributions $|\Psi|^2$,
and formally inhabit the configuration space 
$\cC^N := (\R^d)^N\setminus\bDelta / S_N$
($\bDelta$ denotes the diagonal, and for simplicity we treat the particles as \emph{distinct}), 
which is also the space of $N$-point subsets of $\R^d$.
Conventionally, we may subject $\Psi \in L^2(\R^{dN})$ to a permutation symmetry, 
such as the simple exchange condition for any two particles 
$j \leftrightarrow k$:
\begin{equation} \label{eq:Psi-perm}
	\Psi(\bx_1,\ldots,\bx_k,\ldots,\bx_j,\ldots,\bx_N) 
	= \pm \Psi(\bx_1,\ldots,\bx_j,\ldots,\bx_k,\ldots,\bx_N),
\end{equation}
with $+$ for bosons, and $-$ for fermions,
which is legitimate for $d \ge 3$.
For $d=2$, one may consistently track continuous exchanges of particles and
their winding w.r.t.\ orientation by means of loops in $\cC^N$ modulo continuous
deformations, leaving us with the exchange group $B_N = \pi_1(\cC^N)$, 
the \keyword{braid group on $N$ strands}.
Thus, considering now $\Psi$ as a section of a complex line bundle over $\cC^N$, 
or a $\C$-valued function on the covering space of $\cC^N$,
we may require the equivariance or \keyword{exchange conditions}
\begin{equation} \label{eq:Psi-exch}
	\Psi(\bx_1,\ldots,\bx_k,\ldots,\bx_j,\ldots,\bx_N) 
	= e^{i(2p+1)\alpha\pi} \Psi(\bx_1,\ldots,\bx_j,\ldots,\bx_k,\ldots,\bx_N),
\end{equation}
where $e^{i\alpha\pi} \in \sU(1)$ is the phase of an elementary counter-clockwise exchange, 
with corresponding \keyword{statistics parameter} $\alpha \in \R$ (mod 2).
More generally, if $\Psi$ takes values in $\C^D$ 
(encoding spin or an ``internal'' state), we require
\begin{equation} \label{eq:Psi-exch-op}
	\Psi(\bx_1,\ldots,\bx_k,\ldots,\bx_j,\ldots,\bx_N) 
	= \rho(b_p) \Psi(\bx_1,\ldots,\bx_j,\ldots,\bx_k,\ldots,\bx_N),
\end{equation}
where $\rho\colon B_N \to \sU(D)$ can be any unitary representation
of the braid group, and
$$
	b_p \sim \sigma_1 \sigma_2 \ldots \sigma_p \sigma_{p+1} \sigma_p \ldots \sigma_2 \sigma_1
$$
($\sigma_j$ denoting the generators of $B_N$)
the braid corresponding to the counter-clockwise exchange of two particles
in the case that $p$ other particles are enclosed by the exchange loop.

Here we will only consider the simpler \keyword{abelian anyons} 
\eqref{eq:Psi-exch}
defined by $\C$-valued functions and phases
$\rho(\sigma_j) = e^{i\alpha\pi}$ with fixed $\alpha \in (-1,1]$.
These may be equivalently modeled using bosons (or fermions) with magnetic
interactions, by attaching to each particle a magnetic flux $2\pi\alpha$
(or $2\pi(1-\alpha)$).
In the case of ideal, pointlike particles this will be a point flux of 
Aharonov--Bohm type,
resulting in the typical \keyword{$N$-anyon Hamiltonian}\footnote{We 
choose our units such that the numerical factor on the non-relativistic 
kinetic energy is $\hbar^2/(2m) = 1$.} 
\begin{equation} \label{eq:H-anyon}
	\hH_N = \sum_{j=1}^N \left[ \bigl( -i\nabla_{\bx_j} + \alpha\bA_j \bigr)^2 + V(\bx_j) \right],
	\qquad
	\bA_j = \sum_{k \neq j} \frac{(\bx_j-\bx_k)^\perp}{|\bx_j-\bx_k|^2},
\end{equation}
where we define $(x,y)^\perp := (-y,x)$, 
and $V\colon \R^2 \to \R$ denotes an external potential, 
such as a background formed by other species of particles.
In emergent scenarios, these magnetic point fluxes can be more realistically 
replaced by extended fluxes as well as endowed with other interactions, 
such as Coulomb electrostatic repulsion, or boundary conditions at $\bDelta$
by specification of a suitable operator or quadratic form domain.
We refer to \cite{Lundholm-23} for a recent overview of known properties 
of the many-body ($N \to \infty$) anyon gas, 
including ideal, nonideal, nonabelian and emergent scenarios,
as well as to \cite{LunQva-20} for a mathematical review concerning 
the roles of exchange and exclusion in the ideal abelian and nonabelian anyon gases 
by means of Poincar\'e, Hardy, and Lieb--Thirring inequalities.

\section{Exchange, exclusion, and stability}\label{sec:stability}

The ground-state energy (g.s.e.) of a Hamiltonian such as \eqref{eq:H-anyon}, 
or more generally
\begin{equation} \label{eq:H-general}
	\hH_N = \hT_N + \sum_{j=1}^N V(\bx_j) + \sum_{1 \le j < k \le N} W(\bx_j-\bx_k),
\end{equation}
where $\hT_N$ denotes the kinetic part of the energy and $W$ a symmetric pair interaction,
is defined as the infimum of the corresponding energy functional
$\cE_N[\Psi_N] := \langle \Psi_N, \hH_N \Psi_N\rangle$:
$$
	E_N := \infspec \hat{H}_N = \inf \left\{ \cE_N[\Psi_N] : \norm{\Psi_N}_{L^2} = 1 \right\},
$$
with $\Psi_N$ in the quadratic form domain of $\hH_N$.
We say that the system exhibits \keyword{stability of the first kind} if
$E_N > -\infty$,
which might be true only for certain $N$, such as $N=1$, or for all $N \in \N$.
In the latter case
it exhibits \keyword{stability of the second kind} iff
there exists $C \ge 0$ s.t.\ 
$E_N \ge -CN$, as $N \to \infty$, 
i.e.\ if the energy diverges at most extensively with the number of particles.
Further, note that a stable system may or may not have a ground state 
(consider e.g.\ the case of a free particle in $\R^d$, 
versus the harmonic oscillator potential $V(\bx) = |\bx|^2$).

It turns out that the stability of a realistic many-body system 
is far from trivial to prove mathematically, 
and is not even true for bosons whose energy is non-extensive.
We summarize very briefly a few milestones in the history of the topic
of stability of matter:

\begin{itemize}[wide] 
\item \keyword{Electrogravitics}: 
	An essential component for stability is that 
	the 3D electrostatic Coulomb interaction potential 
	$W(\bx) \propto |\bx|^{-1}$
	admits certain screening and symmetry reduction properties,
	which 
	goes all the way back to Newton \cite{Newton-87}
	in the gravitational context, 
	while the extensivity of a classical Coulomb system with hard-sphere charges 
	was first considered by
	Onsager \cite{Onsager-39}.
	Fisher and Ruelle \cite{FisRue-66} 
	generalized this stability result to extended classical charges.
	Various other mathematically powerful formulations of 
	electrostatic inequalities have been found since,
	such as Baxter's \cite{Baxter-80}.
\item \keyword{Uncertainty principle}: 
	The uncertainty principle of quantum mechanics, usually credited to
	Heisenberg \cite{Heisenberg-25} in its matrix/operator mechanics form, and to
	Schr\"odinger \cite{Schroedinger-26} in its wave mechanics form,
	was found to resolve the classical instability of the hydrogen atom by the 
	quantization of its electron energy levels.
	Fisher and Ruelle used the bound for the lowest eigenvalue
	to prove stability of the first kind for any number of quantum particles $N$.
	More robust functional inequalities were found by
	Hardy \cite{Hardy-20} and
	Sobolev \cite{Sobolev-38},
	that allow to control the unbounded from below potential energy by the kinetic energy.
	In 2D, one may instead employ the inequalities of 
	Ladyzhenskaya \cite{Ladyzhenskaya-58}, 
	Gagliardo \cite{Gagliardo-59}, 
	and Nirenberg \cite{Nirenberg-59}.
	Also the effective functional theories of
	Ginzburg and Landau \cite{GinLan-50,Ginzburg-09} for superconductivity, and
	Gross and Pitaevskii \cite{Gross-61,Pitaevskii-61} for weakly interacting bosons
	(nonlinear Schr\"odinger wave mechanics; see \eqref{eq:GP-func}),
	are subject to stability for small $N$ or weak attractions,
	by means of the uncertainty principle alone.
\item \keyword{Exclusion principle}: 
	Pauli's exclusion principle discovered
	for electrons \cite{Pauli-25}
	is naturally incorporated in any fermionic state \eqref{eq:Psi-perm},
	since $\Psi(\sx) = 0$ at $\sx \in \bDelta$.
	More robustly, such states are spanned by Slater determinants\footnote{%
	For general states this reasoning can be deceiving, however, 
	as illustrated by statistics transmutation in 2D,
	and the actual operator including its domain needs to be considered; 
	see \cite{Forte-91,Lundholm-23}.}
	$$
		\Psi(\bx_1,\ldots,\bx_N) = (u_0 \wedge u_1 \wedge \ldots \wedge u_{N-1})(\bx_1,\ldots,\bx_N),
	$$
	which forbids 
	repeated one-body states/orbitals $u_k \in L^2(\R^d)$, 
	leading to the extensivity of fermionic matter.
	The effective theory of Thomas and Fermi \cite{Thomas-27,Fermi-27} (see \eqref{eq:TF-func}) 
	manifests a degeneracy pressure in a gas of fermions due to exclusion.
	The role of quantum statistics for the stability of the second kind 
	for electrostatic Coulomb systems was finally
	resolved by Dyson and Lenard \cite{DysLen-67,Dyson-67,DysLen-68}. Further,
	Lieb and Simon \cite{LieSim-73,Lieb-76,LieSim-77b} proved stability from the 
	Thomas--Fermi perspective, and subsequently
	Lieb and Thirring \cite{LieThi-75,LieThi-76} found a functional inequality
	for the fermionic kinetic energy in terms of the Thomas--Fermi energy,
	which thereby elegantly combines uncertainty and exclusion.
	Various generalizations of the exclusion principle have also been considered, 
	such as in
	\cite{Gentile-40,Gentile-42,Haldane-91,LunSol-13b,Lundholm-16,LunQva-20}.
\item \keyword{Self-magnetics}: 
	In the case of self-generated magnetic fields, 
	due to their complicated propagation in spacetime 
	(in 3D space; compare our final remarks in Section~\ref{sec:outlook}),
	a fruitful approach has been to minimize the energy over arbitrary fields, 
	i.e.\ to consider
	$$
		E_N := \inf_{\Psi_N,\bA} \cE_N[\Psi_N,\bA],
	$$
	with a functional $\cE_N$
	depending both on an $N$-body quantum state $\Psi_N$
	and an arbitrary magnetic potential $\bA$ 
	that generates a classical magnetic field 
	$\bB = \curl \bA$.
	It is also necessary to include the self-energy of the field $\int_{\R^3} |\bB|^2$.
	For $N=1$ and the Pauli Hamiltonian:
	\begin{equation}\label{eq:Pauli-3d}
		\cE_1[\Psi,\bA] := \int_{\R^3} \Bigl( 
			\bigl|({-i}\nabla + \sqrt{a}\bA)\Psi\bigr|^2 
			+ \sqrt{a} \langle\Psi,\bB \cdot \boldsymbol{\sigma}\Psi\rangle 
			- \frac{Za}{|\bx|} |\Psi|^2 + \frac{1}{4\pi} |\bB|^2
			\Bigr),
	\end{equation}
	$\Psi \in H^1(\R^3;\C^2)$, 
	$\boldsymbol{\sigma} = (\sigma_1,\sigma_2,\sigma_3)$ the Pauli matrices,
	and it was shown by Fr\"ohlich, Lieb and Loss \cite{FroLieLos-86} 
	that stability holds iff $Za^2$ is small enough,
	although the result is complicated by the existence of zero modes for the Pauli operator,
	exemplified by Loss and Yau \cite{LosYau-86}.
	Stability of the second kind for $N \to \infty$ and self-generated fields in 3D
	was finally proved by
	Fefferman \cite{Fefferman-95} and
	Lieb, Loss and Solovej \cite{LieLosSol-95},
	assuming that both the maximal charge of nuclei $Z$ 
	and the fine-structure constant $a$ 
	are (realistically) small enough.
\end{itemize}

Much more work has been done on stability and much more can be said about it, 
such as relativity, quantized electromagnetic fields, etc., 
and we refer to the textbook of Lieb and Seiringer \cite{LieSei-10} 
for a comprehensive 
treatment of the topic, 
as well as to \cite{Lundholm-17} for a few complementary aspects 
such as a more recent, local approach to uncertainty and exclusion suitable for anyons,
which is rooted in the Poincar\'e \cite{Poincare-90} and Hardy inequalities 
(cp.\ \cite{HofLapTid-08}).
Most relevantly,
a Lieb--Thirring inequality for ideal anyons with the Hamiltonian \eqref{eq:H-anyon}
has been proved by the author and Solovej \cite{LunSol-13a}, 
with subsequent refinements in \cite{LarLun-16,LunSei-17,LunQva-20}:
\begin{equation}\label{eq:LT-anyon}
	\langle \Psi, \hH_N \Psi \rangle
	\ge \int_{\R^2} \Bigl( \CLT(\alpha) \,\varrho_\Psi^2 + V\varrho_\Psi \Bigr)
	\qquad \forall \Psi \in H^1_\sym(\R^{2N};\C),
\end{equation}
where $\varrho_{\Psi}$ is the one-body density associated to $\Psi$,
normalized $\int_{\R^2} \varrho_\Psi = N$,
and a constant
$$
	|\alpha| \lesssim \CLT(\alpha) \lesssim |\alpha|, \qquad \alpha \in [-1,1].
$$
Thereby follows stability of the second kind for a system of $N$ ideal anyons 
in the plane with a background of other particles and 3D Coulomb interactions
(i.e.\ confined particles but unconfined electrostatics),
as long as $\alpha \notin 2\Z$; see \cite[Thm.~21]{LunSol-14}, \cite{Lundholm-17}.

\section{A density functional theory for almost-bosonic anyons}\label{sec:DFT}

For bosonic and fermionic systems, various 
effective energy functionals have been proposed 
in order to describe the many-body ground states 
and their corresponding probability densities quantitatively, 
the most famous being (here in their 2D formulations, with energy per particle,
i.e.\ limits of $E_N/N$ as $N \to \infty$):
\begin{itemize}
\item
The \keyword{Gross--Pitaevskii (GP) functional} for weakly interacting bosons:
\begin{equation}\label{eq:GP-func}
	\cE^{\mathrm{GP}}[u] := \int_{\R^2} \Bigl( \bigl|({-i}\nabla + \bA_{\rm ext})u\bigr|^2 + g |u|^4 + V|u|^2 \Bigr),
	\qquad \int_{\R^2} |u|^2 = 1,
\end{equation}
where $u \in L^2(\R^2)$ is the one-body state into which the Bose gas condenses,
and $g \in \R$ an effective coupling strength of a scalar interaction.

\item
The \keyword{Thomas--Fermi (TF) functional}:  
\begin{equation}\label{eq:TF-func}
	\cETF[\varrho] :=
	\int_{\R^2} \Bigl( \CTF N \varrho^2 + V\varrho \Bigr),
	\qquad \int_{\R^2} \varrho = 1,
\end{equation}
which formally arises both as a limit of the GP functional
as the coupling $g$ becomes strong, with $\varrho = |u|^2$, 
as well as for $N$ non-interacting fermions, 
with their one-body density $\varrho = \varrho_\Psi \ge 0$ 
suitably
normalized 
and the constant given by the energy per particle and unit 
density of the homogeneous free Fermi gas, which is $C^{\mathrm{TF}} = 2\pi$ in 2D.
\end{itemize}

In the case of anyons, an \keyword{``average-field'' approximation} 
(or perhaps more appropriately, ``constant-field'')
was initially proposed as a simple interpolation between bosons and fermions 
for $0 \le \alpha \le 1$:
\begin{equation}\label{eq:cf-func}
	\cE^{\mathrm{af}}[\varrho] \approx
	\int_{\R^2} \Bigl( 2\pi \alpha N \varrho^2 + V\varrho \Bigr),
\end{equation}
i.e.\ TF with the constant $\CTF(\alpha) = 2\pi\alpha$,
which is motivated by 
estimating locally at each point $\bx \in \R^2$ 
the ground-state energy $E_n = |B|n$ of $n \approx N\varrho(\bx)d\bx$ 
bosons in the lowest Landau level of a constant
magnetic field with strength $B \approx 2\pi\alpha N\varrho(\bx)$
(a local density approximation).
However, for anyons close to bosons one should better consider the following functional:
\begin{equation}\label{eq:AFP-func}
	\boxed{
	\cE_{\beta,\gamma,V}[u] := \int_{\R^2} \left[
		\left| (-i\nabla + \beta\bA[|u|^2]) u\right|^2
		+ \gamma |u|^4 + V|u|^2 \right], 
	}
\end{equation}
where:
\begin{itemize}
\item $u \in H^1(\R^2;\C)$ is a one-body quantum state with probability density 
	$\varrho = |u|^2$, subject to the normalization
	$\int_{\R^2} \varrho = 1$;

\item $\bA[\varrho] \colon \R^2 \to \R^2$ is a magnetic vector potential
	which generates a magnetic field proportional to $\varrho$:
	\begin{equation} \label{eq:mag-pot}
		\bA[\varrho](\bx) := (\nabla^\perp w_0) * \varrho (\bx) 
		= \int_{\R^2} \frac{(\bx-\by)^\perp}{|\bx-\by|^2} \varrho(\by) d\by,
		\qquad w_0(\bx) := \log |\bx|,
	\end{equation}
	so that
	\begin{equation} \label{eq:self-field}
		\curl \beta\bA[\varrho](\bx) = \beta (\Delta w_0) * \varrho (\bx) = 2\pi\beta\varrho(\bx),
	\end{equation}
	where we used that $\nabla^\perp \cdot \nabla^\perp = \Delta$ 
	and that $w_0$ is the fundamental solution in 2D:
	$$
		\Delta w_0 = 2\pi \delta_0;
	$$
	
\item $\beta \in \R$ is the strength of this magnetic self-interaction
	(the total/fractional number of flux units of the field,
	i.e. $\beta \sim \alpha N$ for anyons with $\alpha$ flux units per particle);
	
\item $\gamma \in \R$ is the strength of a scalar self-interaction,
	being attractive if $\gamma < 0$ and repulsive if $\gamma > 0$;

\item
	the external potential $V\colon \R^2 \to \R$ is usually assumed bounded from below and trapping,
	i.e. $V(\bx) \to +\infty$ as $|\bx| \to \infty$.
	As in the GP and TF theories, 
	one may also add 
	an external magnetic field potential $\bA_{\rm ext}$.
\end{itemize}
Thus, we define the ground-state energy (per particle)
\begin{equation}\label{eq:AFP-gse}
	E_{\beta,\gamma,V} := 
		\inf \left\{ \cE_{\beta,\gamma,V}[u] : 
		u \in H^1(\R^2;\C), 
		\int_{\R^2} |V| |u|^2 < \infty,
		\int_{\R^2} |u|^2 = 1 \right\}.
\end{equation}

\begin{figure}
\includegraphics[scale=0.5]{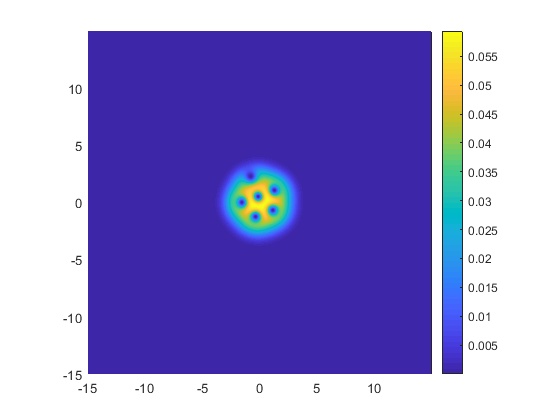}
\includegraphics[scale=0.5]{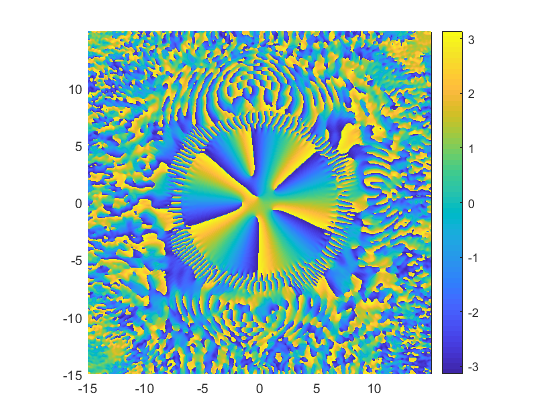}
\caption{Density and phase of an approximate minimizer of 
	$\cE_{\beta,\gamma,V}$ at $\beta=10$, $\gamma=20\pi$, $V=x^2+y^2$, 
	energy $\approx 9.066$.}
\label{fig:beta10}
\end{figure}

\begin{figure}
\includegraphics[scale=0.5]{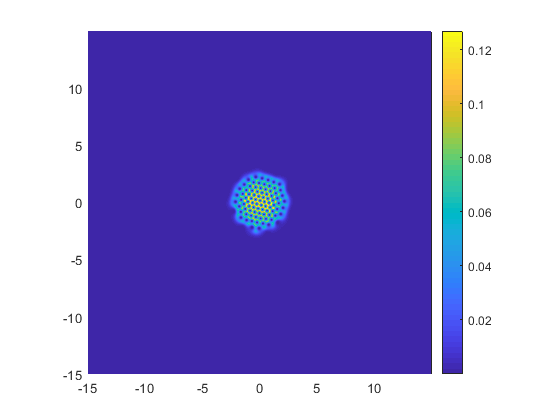}
\includegraphics[scale=0.5]{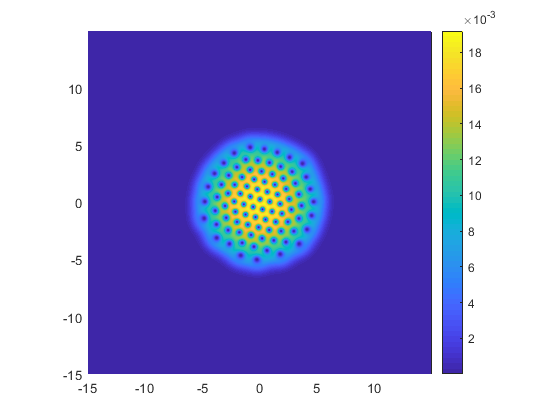}
\caption{Density of an approximate minimizer of $\cE_{\beta,\gamma,V}$ 
	at $V=x^2+y^2$, $\beta=100$, and $\gamma=-186\pi$ (left), 
	respectively $\gamma=200\pi$ (right).}
\label{fig:beta100}
\end{figure}

It was shown in \cite{LunRou-15}
that the minimizers $u^{\mathrm{af}}$
of the 
problem \eqref{eq:AFP-gse}
at $\gamma=0$
describe the condensed ground states $\Psi \approx \otimes^N u^{\mathrm{af}}$ of a nonideal anyon gas
in a bosonic limit $N \to \infty$, $\alpha = \beta/N \to 0$, $\beta$ fixed,
if the fluxes remain sufficiently extended
while taking their extension slowly to zero, then
corresponding to a weak but dense self-generated magnetic field.
This mean-field (or ``average-field''; cp.\ \cite{Wilczek-90}) 
result was strengthened and generalized in \cite{Girardot-20,Girardot-21} 
to include external magnetic fields.
In \cite{CorLunRou-17,CorLunRou-proc-17}
it was shown that the average (mesoscopic) density profile $\varrho$ of the
minimizers of the functional $\cE_{\beta,\gamma,V}$ for $\gamma=0$ and $\beta \gg 1$
can be described using the minimizing density of a TF functional 
\eqref{eq:TF-func} 
with a constant
\begin{equation} \label{eq:af-TF-const}
	\CTF(\beta) = 2\pi c\beta, \qquad
	c \approx 2\sqrt{\pi}/3 \approx 1.18,
\end{equation}
thus modifying the naive constant-field approximation \eqref{eq:cf-func} 
with a slightly higher energy.
Numerical methods were used to estimate this universal constant $c$ and to investigate
the ground states, which, as $\beta$ grows large, manifest the emergence of a 
triangular vortex lattice formed of approximately $\beta$ 
singly-quantized vortices and
with a scale set by the density profile \cite{CorDubLunRou-19}.
The microscopic inhomogeneity of the lattice explains the factor 
\eqref{eq:af-TF-const} of energy increase 
(greater than the Abrikosov factor of rotating condensates; cp.\ \cite{AftBlaNie-06b})
compared to the constant self-generated field for bosons 
and the homogeneous Fermi gas.
Further, it has been proposed in the physics literature \cite{Sen-91,SenChi-92,ChiSen-92}
that for ideal anyons close to bosons it is more appropriate to include 
a scalar self-energy term with $\gamma = 2\pi|\beta|$.
Indeed, we expect that different interactions or types of anyons can 
can be effectively described by the functional 
\eqref{eq:AFP-func} with different $\gamma$, both positive and negative
(see \cite{Nguyen-24} for initial results).
Some examples of approximate minimizers of \eqref{eq:AFP-func},
found by numerical optimization,
for various $\beta$ and $\gamma$ are given in Figures~\ref{fig:beta10} and \ref{fig:beta100}.

The Euler--Lagrange equation for a ground state $u$ of \eqref{eq:AFP-gse} 
with finite energy is
\begin{equation}\label{eq:EL}
\Big[-\left(\nabla + i\beta\bA[|u|^2]\right)^2 
- 2\beta \bA * \bigl[\beta\bA[|u|^2]|u|^2 + \bJ[u]\bigr] 
+ 2\gamma|u|^2 + V\Big] u = \lambda u,
\end{equation}
where
	$\bJ[u] := \frac{i}{2}(u\nabla \bar{u} - \bar{u}\nabla u)$
denotes the current of $u$, and $\lambda = \lambda(u) \in \R$ is a constant;
see \cite[Appendix]{CorLunRou-17} and \cite[Rmk.~3.11]{AtaLunNgu-24}.
For $\beta=0$ it reduces to the \emph{local}, nonlinear Schr\"odinger (NLS) equation 
associated to the GP (NLS) functional \eqref{eq:GP-func}.
For $\beta \neq 0$ the \emph{nonlocal}, nonlinear equation and energy functional 
arise in the \keyword{Chern--Simons--Schr\"odinger (CSS)}
or \keyword{Ginzburg--Landau--Higgs (GLH)} theory, 
in which the magnetic field \eqref{eq:self-field} is self-consistently coupled to the density; 
see, e.g., \cite{RebSol-84,Dunne-95,Dunne-99,Khare-05,Tarantello-08}.
Therefore the states $u$ are also called CSS wave functions and \eqref{eq:AFP-func} 
is also called the CSS or \keyword{``average-field-Pauli'' (afP)} functional
(cp.\ the 3D Pauli operator in \eqref{eq:Pauli-3d}).

\section{Magnetic stability and nonlinear Landau levels}\label{sec:magstab}

Let us define the \keyword{critical coupling} as the
minimal quotient of magnetic kinetic to scalar (TF-type) self-energy:
\begin{equation} \label{eq:defgamma}
		\gamma_*(\beta) := \inf \left\{ 
			\frac{ \cE_{\beta,0,0}[u]}{\int_{\R^2} |u|^4}
		: u \in H^1(\R^2;\C), \int_{\R^2} |u|^2 = 1 \right\}.
\end{equation}
We also define the \keyword{nonlinear Landau level (NLL)} for any $\beta \neq 0$:
\begin{equation} \label{eq:defNLL}
		\NLL(\beta) := \left\{ u \in H^1(\R^2;\C) :
			\cE_{\beta,0,0}[u] = 2\pi|\beta|\int_{\R^2} |u|^4,
			\int_{\R^2} |u|^2 = 1 \right\}.
\end{equation}
Our main result for the stability of the system  
\eqref{eq:AFP-func}-\eqref{eq:AFP-gse} may then be summarized as follows:

\begin{theorem}[Stability for the almost-bosonic anyon gas; {\cite[Thm.~3]{AtaLunNgu-24}}]\label{thm:stability}
    Let $\beta \in \R$ and the potential $V$ be smooth and bounded from below. 
    The critical coupling $\gamma_*(\beta)$ defined in \eqref{eq:defgamma}
    is exactly the critical value for stability of 
    the system \eqref{eq:AFP-gse} in the sense that 
    $E_{\beta,\gamma,V} > -\infty$ if $\gamma \ge -\gamma_*(\beta)$
    and $E_{\beta,\gamma,V} = -\infty$ if $\gamma < -\gamma_*(\beta)$.
    If $V=0$ 
    then $E_{\beta,-\gamma_*(\beta),0}=0$ for all $\beta$, 
    and if furthermore
    $\beta\ge 2$ and $-\gamma=\gamma_*(\beta)$, 
    which in this case equals $2\pi\beta$,
    then zero-energy ground states exist if and only if $\beta \in 2\N$ 
    and are then given exactly by the $2\beta$-dimensional soliton manifold 
	\begin{multline}\label{eq:NLL}
		\NLL(\beta=2n) = \Biggl\{ u = \frac{1}{\sqrt{\pi n}} \, \frac{\overline{P' Q - P Q'}}{|P|^2 + |Q|^2} : \ 
			\text{$P,Q$ coprime and linearly independent}\\
			\text{complex polynomials s.t. $\max\{\deg P,\deg Q\}=n$} 
			\Biggr\},
	\end{multline}
	whereas $\NLL(\beta) = \emptyset$ if $\beta \notin 2\Z$.
	Finally, for any $\beta,\gamma \in \R$,
	\begin{equation}\label{eq:AFP-KLT}
		E_{\beta,\gamma,V} \ge E^{\rm KLT}_{\beta,\gamma,V} 
		:= \inf \left\{
		\int_{\R^2} \left[ (\gamma_*(\beta)+\gamma) \varrho^2 + V\varrho \right]
		: \varrho \in L^2(\R^2;\R_+), \ \int_{\R^2} \varrho = 1
		\right\}.
	\end{equation}
\end{theorem}

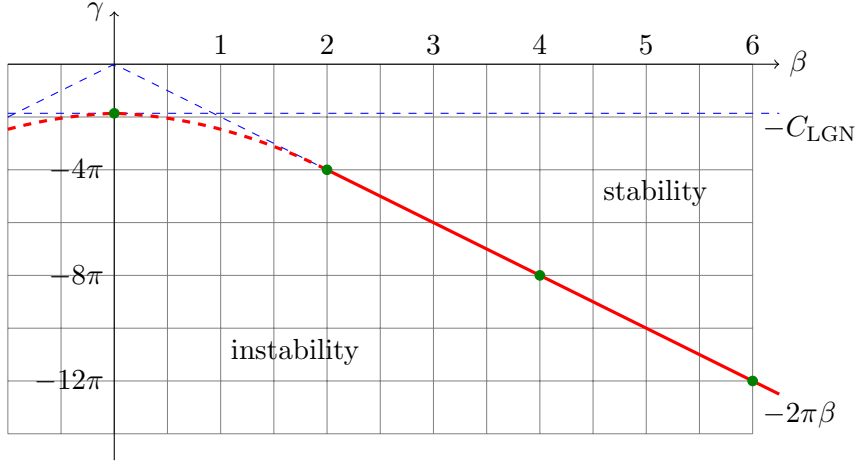
\begin{figure}
    \begin{tikzpicture}[domain=-2:12,scale=0.7]
	    \draw[very thin,color=gray] (-2,-7) grid (12,0);
	    \draw[->] (-2,0) -- (12.5,0) node[right] {$\beta$};
	    \draw[->] (0,-7.5) -- (0,1) node[left] {$\gamma$};
	    \draw[thin,dashed,color=blue,domain=-2:12.5,samples=50] plot (\x,{-0.93});
	    \draw[thin,dashed,color=blue,domain=-2:4,samples=50] plot (\x,{-0.5*abs(\x)});
	    \draw[very thick,dashed,color=red,domain=-2:4,samples=50] plot (\x,{-0.5*abs(\x) - 0.93*(4-abs(\x))^2/16});
	    \draw[very thick,color=red,domain=4:12.5,samples=50] plot (\x,-\x/2);
		\draw [fill,color=darkgreen] (0,-0.93) circle [radius=0.09];
		\draw [fill,color=darkgreen] (4,-2) circle [radius=0.09];
		\draw [fill,color=darkgreen] (8,-4) circle [radius=0.09];
		\draw [fill,color=darkgreen] (12,-6) circle [radius=0.09];
	    \node[above] at (2,0) {$1$};
	    \node[above] at (4,0) {$2$};
	    \node[above] at (6,0) {$3$};
	    \node[above] at (8,0) {$4$};
	    \node[above] at (10,0) {$5$};
	    \node[above] at (12,0) {$6$};
	    \node[left] at (0,-2) {$-4\pi$};
	    \node[left] at (0,-4) {$-8\pi$};
	    \node[left] at (0,-6) {$-12\pi$};
	    \node[below right] at (12,-0.8) {$-\CLGN$};
	    \node[below right] at (12,-6.2) {$-2\pi\beta$};
		\node [below right] at (9,-2) {stability};
		\node [below right] at (2,-5) {instability};
    \end{tikzpicture}
\caption{Sketch of $-\gamma_*(\beta)$ with exact $\NLL$s at $\beta = 2n$.}
\label{fig:stability}
\end{figure}

	We also note the following:
\begin{itemize}
\item by \keyword{complex conjugation (orientation) symmetry}, 
	$u \mapsto \overline{u}$, $\beta \mapsto -\beta$,
	we have that $\gamma_*(-\beta) = \gamma_*(\beta)$ 
	and $\NLL(-\beta) = \overline{\NLL(\beta)}$ 
	for all $\beta \in \R \setminus \{0\}$.
	Therefore we may restrict to $\beta \ge 0$, without loss of generality.

\item by the \keyword{diamagnetic inequality} 
	(loosely speaking, writing $u = |u|e^{i\varphi}$ 
	and neglecting the nonnegative phase contribution; 
	see, e.g., \cite[Thm.~7.21]{LieLos-01}):
	$$
		\cE_{\beta,0,0}[u] \ge \int_{\R^2} \bigl|\nabla|u|\bigr|^2 
		\qquad \forall \beta \in \R, \ u \in H^1,
	$$
	so that $\gamma_*(\beta) \ge \gamma_*(0) =: \CLGN$,
	defined as the optimal constant of the embedding $H^1 \hookrightarrow L^4$ due to  
	Ladyzhenskaya--Gagliardo--Nirenberg (LGN) 
	\cite{Ladyzhenskaya-58,Gagliardo-59,Nirenberg-59}:
	\begin{equation}\label{eq:LGN}
		\int_{\R^2} |\nabla u|^2 \int_{\R^2} |u|^2 \ge \CLGN \int_{\R^2} |u|^4
		\qquad \forall u \in H^1.
	\end{equation}
	Further, by Weinstein \cite{Weinstein-83}, 
	$\CLGN = \int_{\R^2} |\tau|^2 /2 \approx 0.931 \times 2\pi$,
	where $\tau \in H^1 \cap C^\infty(\R^2;\R^+)$,
	known as ``\keyword{Townes soliton}'' \cite{ChiGarTow-64,ChiGarTow-65,Fibich-15}, 
	is the unique minimizer for $\gamma_*(0)$ that saturates \eqref{eq:LGN},
	up to translations, rescaling, and a constant phase.
	
\item by a \keyword{supersymmetric Pauli/Bogomolnyi bound} 
	(cf. \cite{Bogomolny-76,HloSpe-93}, and \eqref{eq:susy-trick}):
	$$
		\int_{\R^2} \bigl|(\nabla + i\bA)u\bigr|^2 \ge \pm \int_{\R^2} B |u|^2,
		\qquad B = \curl \bA,
	$$
	where for the self-generated field we use \eqref{eq:self-field},
	so that $\gamma_*(\beta) \ge 2\pi|\beta|$ for all $\beta \in \R$.
	The corresponding coupling $\gamma = \pm 2\pi\beta$ is known as 
	the \keyword{self-dual coupling} in the literature,
	and the corresponding energy equality \eqref{eq:defNLL} satisfied by $u \in \NLL$ 
	is known as a \keyword{Pohozaev-Bogomolnyi identity}.
\end{itemize}
We may view the CSS wave function
$u\colon \R^2 \to \C$ as a global section of a $\sU(1)$ complex line bundle
over $\R^2 \cong \C$ with self-generated curvature $B = 2\pi\beta|u|^2$,
and thus minimizing \eqref{eq:defgamma} amounts to finding the optimal 
embedding of such a section subject to its own curvature.

\smallskip

The first part of Theorem~\ref{thm:stability} follows straightforwardly by
taking dilations of $u \in H^1$,
\begin{equation}\label{eq:scaling}
u_{\lambda}(\bx) := \lambda u(\lambda \bx),
\quad \lambda > 0,
\end{equation}
which preserves the $L^2$-norm but scales both the magnetic self-energy $\cE_{\beta,0,0}$ 
and the $L^4$-term by $\lambda^2$ (see \cite[Lemma~3.4]{CorLunRou-17}).
Also, the bound \eqref{eq:AFP-KLT}, 
or \keyword{Keller--Lieb--Thirring inequality}, 
is an immediate consequence of the definition \eqref{eq:defgamma}.
However, the nontrivial core of the result is the following theorem which describes the main behavior
of the critical coupling \eqref{eq:defgamma} and of the corresponding ground states:

\begin{theorem}[Magnetic stability; {\cite[Thm.~2]{AtaLunNgu-24}}] \label{thm:magneticstability}
The following holds:
\begin{enumerate}[label=\text{(\roman*)}]
\item\label{itm:mstab-gamma}
	We have that $\beta \mapsto \gamma_*(\beta)$ is a Lipschitz function and satisfies
	\begin{equation}\label{eq:mag-bermuda}
        \gamma_*(\beta) > \max\{\CLGN, 2\pi\beta\}
        \quad \text{for every $0<\beta <2$,}
	\end{equation}
    and 
	\begin{equation}\label{eq:mag-susy}
        \gamma_*(\beta) = 2\pi\beta 
        \quad \text{for every $\beta \geq 2$.}
	\end{equation}
\item\label{itm:mstab-mini}
	Any minimizer of \eqref{eq:defgamma}, if it exists, is smooth. For small enough $0 < \beta < 2$, there exists a minimizer. For $\beta \geq 2$,
    minimizers exist if and only if $\beta \in 2 \N$, 
	and are of the form 
	\begin{equation}\label{eq:mag-solution}
		u = u_{P,Q} := \sqrt{\frac{2 }{\pi \beta}} \, \frac{\overline{P' Q - P Q'}}{|P|^2 + |Q|^2},
	\end{equation}
	where $P,Q$ are two coprime and linearly independent complex polynomials satisfying 
	$$
		\max (\deg(P),\deg(Q)) = \frac{\beta}{2}.
	$$
\item\label{itm:mstab-symm}
	Finally, $u_{P,Q} = u_{\tilde{P},\tilde{Q}}$ for two such pairs of polynomials $(P,Q),(\tilde{P},\tilde{Q})$ if and only if
	$(P,Q)=\Lambda(\tilde{P},\tilde{Q})$ for some constant $\Lambda \in \R^+ \times \sSU(2)$.
\end{enumerate}
\end{theorem}

The proof of this theorem uses an Aharonov--Casher-type factorization result of 
supersymmetric quantum mechanics 
(cf. \cite{AhaCas-79,Jackiw-84,Jackiw-86,ErdVou-02}):
\begin{equation}\label{eq:susy-trick}
	\int_{\R^2} \left[ |(\nabla + i\bA)u|^2 \pm B |u|^2 \right]
	= \int_{\R^2} \bigl|(\partial_1 \pm {\rm i}\partial_2)(e^{\pm \psi/2}u)\bigr|^2 e^{\mp \psi},
	\quad \text{if $\textstyle\bA = -\frac{1}{2}\nabla^\perp \psi$},
\end{equation}
where $\psi$ is a corresponding \keyword{superpotential},
together with the exact solution of a generalized Liouville equation.
Namely, with $\beta > 0$, the self-dual coupling $\gamma=-2\pi\beta$, 
and the minus sign on $B$,
the kernel of \eqref{eq:susy-trick} is given by
$$
	u(\bx) = (4\pi\beta)^{-1/2} e^{\psi(\bx)/2} \overline{f(z)}
	\qquad \Rightarrow \qquad 
	4\pi\beta \varrho = e^{\psi} |f|^2,
$$
where $f\colon \C \to \C$ is analytic, $\partial_z = \frac{1}{2}(\partial_1 - i\partial_2)$. 
Further, if $f$ has a limited growth as $z \to \infty$, 
i.e. $f(z) = C\prod_{j=1}^M (z-z_j)$ is a polynomial, 
then the normalized density $\varrho = |u|^2$, 
respectively superpotential $\psi$, must satisfy
$$
	2\pi\beta\varrho = B = \curl \bA = -\frac{1}{2} \Delta \psi,
$$
i.e.
\begin{equation}\label{eq:Liouville-psi}
	\boxed{-\Delta \psi = |f|^2 e^{\psi}}
	\qquad \Leftrightarrow \qquad
	{-\Delta} \log(\varrho) = 4\pi\beta\varrho - 4\pi \sum_{j=1}^M \delta_{z_j}.
\end{equation}
The solutions of this equation are given exactly by the densities of \eqref{eq:mag-solution},
and require $\beta \in 2\N$:

\begin{theorem}[Generalized Liouville equation; {\cite[Thm.~1]{AtaLunNgu-24}}]
	Let $f\colon \C \to \C$ 
	be a nonzero polynomial.
	All the weak solutions $\psi \in L^1_\loc(\R^2;\R)$
	of \eqref{eq:Liouville-psi}, 
	such that $\int_{\R^2} |f|^2 e^{\psi} < \infty$,
	are of the form
	$$
		\psi = \psi_{P,Q} := \log(8) - 2\log(|P|^2 + |Q|^2),
	$$
	where $P,Q$ are two coprime complex polynomials which satisfy $f= P' Q -P Q'$. 
	Moreover,
	$\max(\deg(P),\deg(Q)) = \frac{\int_{\R^2}|f|^2 e^{\psi} }{8\pi} = \frac{\beta}{2}$,
	and if $(P,Q)$ and $(\tilde{P},\tilde{Q})$
	are pairs of polynomials (not necessarily coprime) then
	$\psi_{P,Q} = \psi_{\tilde{P},\tilde{Q}}$ if and only if
	$(\tilde{P},\tilde{Q}) = \Lambda(P,Q)$, 
	for some constant $\Lambda \in \sU(2)$.
\end{theorem}

The necessary regularity for the above conclusions is derived from the Euler--Lagrange equation \eqref{eq:EL},
and furthermore a concentration-compactness approach is used to obtain 
the result \eqref{eq:mag-bermuda} on the coupling, which can be interpreted as
a \keyword{breaking of supersymmetry} at $|\beta| < 2$.
It is further noted that the number $M$ of zeros $z_j$ of the Wronskian $f = P'Q-PQ'$ satisfies
$$\beta/2-1 \le M \le \beta-2,$$
which is also interpreted as the \keyword{vorticity} of the state $u_{P,Q} \in \NLL(\beta)$.
Hence, the vorticity of CSS ground states at critical, self-dual 
coupling increases linearly with $\beta$.

\medskip

As examples we have the following radially symmetric, exact soliton solutions:
\begin{itemize}
	\item ``Townes soliton'' at $\beta=0$:
	$u \sim \tau$ the unique positive and radial solution 
	to the critical self-focusing 2D nonlinear Schr\"odinger equation
	$$
		-\Delta u - |u|^2 u = -u
	$$
	at the critical coupling 
	$\gamma_*(0) = \CLGN = \norm{\tau}_{L^2}^2/2 \approx 0.931 \times 2\pi$: 
	$$
		\int_{\R^2} |\nabla \tau|^2 \int_{\R^2} |\tau|^2 = \CLGN \int_{\R^2} |\tau|^4.
	$$
	No analytical expression is known for $\tau$. 
	Its shape is roughly similar to a Gaussian but with the asymptotic behavior
	(see \cite[Fig.~3.2 \& Lem.~ 6.14]{Fibich-15})
	$$
		\tau(\bx) \sim C r^{-1/2}e^{-r},
		\qquad r = |\bx| \gg 1.
	$$
	
	\item ``{\it versiera}'' or ``the witch of Agnesi'' 
	(see \cite[p.~178]{Struik-69}) at $\beta=2$:
	$$
		u(z) = \frac{1}{\sqrt{\pi}} \frac{1}{|z|^2+1},
		\qquad \text{(ex. $P(z)=z$, $Q(z)=1$)}
	$$
	
	\item ``vortex ring'' at $\beta=2n$, $n>1$:
	$$
		u(z) = \sqrt{\frac{n}{\pi}} \frac{\bar{z}^{n-1}}{|z|^{2n}+1},
		\qquad \text{(ex. $P(z)=z^n$, $Q(z)=1$)}
	$$
\end{itemize}
The first one was found already in \cite{ChiGarTow-64} in a geometric optics context
(see also \cite{ChiGarTow-65,Fibich-15}), 
and the others by Jackiw and Pi in \cite{JacPi-90b} by stretching the applicability of 
Liouville's solution for the regular Liouville equation \cite{Liouville-53}.
They later also found a more general formula 
for the soliton states in $\NLL(\beta=2n)$ corresponding to distinct roots,
and thereby conjectured that $\dim \NLL(\beta=2n) = 4n$ \cite{JacPi-90a}.
The mathematical precision of these results was discussed in 
\cite{Hagen-91,JacPi-91b,HorYer-98}.
See also \cite{Dunne-99,HorZha-09} for more recent reviews,
and \cite{Eremenko-21,LiLiu-22,AtaLunNgu-24,Ataei-24} 
for recent mathematical developments.

\section{Conclusions and outlook}\label{sec:outlook}

We summarize the main points:
\begin{itemize}
	\item In 2D, intermediate {\bf exchange} quantum statistics is possible, 
	with simple exchange phase $e^{i\alpha\pi}$, resulting in ``anyons''.
	\item For $\alpha \notin \Z$ and $N \gg 1$, the relationship between {\bf exchange} 
	and {\bf exclusion} is a difficult mathematical problem,
	although {\bf stability} is known for ideal abelian anyons.
	\item A precise DFT has been identified in an {\bf almost-bosonic} limit 
	$\alpha \to 0$, $N \to \infty$, s.t. $\alpha N \to \beta \in \R$;
	also including scalar self interactions with coupling strength $\gamma \in \R$.
	\item There is linearly increasing stability in this model for $|\beta| \ge 2$: 
	$\boxed{\gamma \ge -2\pi|\beta|,}$
	i.e.\ {\bf extensivity} w.r.t.\ the total flux.
	\item For $|\beta| \ge 2$, stability at the critical coupling $\gamma = -2\pi|\beta|$
	is saturated exactly at $\beta \in 2\Z$ by a manifold of soliton states of dimension $2|\beta|$,
	which we call ``{\bf nonlinear Landau levels}''.
\end{itemize}

Some ongoing work includes:
\begin{itemize}
	\item The derivation of the effective functional \eqref{eq:AFP-func} from the underlying many-body quantum mechanics at $\gamma \neq 0$.
	See \cite{Nguyen-24} for some initial results, including a refined analysis 
	of the collapse phenomenon at $\beta \to 0$.
	\item External fields, potentials, and a generally better understanding through numerics.
	\item Fractional flux $\beta/2$, and replacing global functions by local sections.
\end{itemize}

Finally, one may ask why we care so much about Flatland, a $2+1$-dimensional world 
which might seem artificial. 
However, let us point out that such scenarios are nowadays routinely realized 
in the lab by confinement using strong potentials and magnetic fields, 
and recently with strong evidence for emergent anyons 
\cite{Bartolomei-etal-20,Nakamura-etal-20}.
Further, 
anyons are also relevant from a fundamental physics perspective, 
both as a toy model for quantized gravity on a locally flat plane 
\cite{tHooft-88,DesJac-88}, 
and possibly also as a holographic realization of actual $3+1$-dimensional 
quantum gravity and cosmology; compare, e.g., \cite{PitRui-15}.
Thus, we may view this topic as a natural continuation of 
the research program of Jackiw:

\smallskip

{\it ``It is the purpose of our research program to study in three-dimensional space-time the classical and quantum motions of matter that interacts gravitationally. Since there are no propagating gravitational degrees of freedom, the problem is tractable, and we can learn much about the puzzles that are encountered when a geometrical theory is confronted by quantum mechanics. In four dimensions these puzzles exist as well, and it is my opinion that understanding them is important for understanding quantum gravity; a task quite independent of and perhaps more fundamental than the task of overcoming the unrenormalizable infinities that pollute four-dimensional gravity, but are absent in three dimensions since non-renormalizable graviton exchange does not occur.''}

Roman Jackiw, in {\it Topics in planar physics}, 1990 \cite{Jackiw-90}.

\medskip\noindent\textbf{Acknowledgments.} 
A warmest thanks is due to the organizers of ICGTMP, Group33/35, 
who persisted through the pandemic and created such a wonderful gathering
from all corners of our globe. 
This contribution is based primarily on the overview article \cite{Lundholm-23} 
and the recent work \cite{AtaLunNgu-24}. 
The author wishes to express his thanks to 
Alireza Ataei and Dinh-Thi Nguyen for fruitful collaboration on that work,
as well as to Michele Correggi and Nicolas Rougerie for
collaborations on earlier works upon which the present one rests,
and last but not least to Romain Duboscq who supplied the Matlab code 
from which much insight has been gained.
Financial support from the Swedish Research Council 
(grant no.\ 2021-05328, ``Mathematics of anyons and intermediate quantum statistics'') 
is gratefully acknowledged.


\def\MR#1{} 

\newcommand{\etalchar}[1]{$^{#1}$}


\end{document}